\documentclass[12pt]{article}
\usepackage{amsmath}
\usepackage[dvips]{graphicx}

\begin{document}

\begin{center}
\textbf{\large On the Tensor/Scalar Ratio in Inflation with UV Cutoff}\\[0pt%
]
\vspace{48pt}

A. Ashoorioon\footnote{amjad@astro.uwaterloo.ca}, R.B. Mann\footnote{%
mann@avatar.uwaterloo.ca}

\vspace{12pt}

\textit{Departments of Physics, University of Waterloo,} \\[0pt]
\textit{\ Waterloo, Ontario, N2L 3G1, Canada\\[0pt]}

\begin{abstract}
\addtolength{\baselineskip}{1.2mm}
\addtolength{\baselineskip}{1.2mm} Anisotropy of the cosmic
microwave background radiation (CMB) originates from both tensor
and scalar perturbations.  To study the characteristics of each of
these two kinds of perturbations, one has to determine the
contribution of each to the anisotropy of CMB. For example, the
ratio of the power spectra of tensor/scalar perturbations can be
used to tighten bounds on the scalar spectral index. We
investigate here the implications for the tensor/scalar ratio of
the recent discovery (noted in astro-ph/0410139) that the
introduction of a minimal length cutoff in the structure of
spacetime does not leave boundary terms invariant.  Such a cutoff
introduces an ambiguity in the choice of action  for tensor and
scalar perturbations, which in turn can affect this ratio. We
numerically solve for both tensor and scalar mode equations in a
near-de-sitter background and explicitly find the cutoff
dependence of the tensor/scalar ratio during inflation.
\end{abstract}

\end{center}

\begin{equation*}
\end{equation*}

\hfill{} \vspace{32pt}

\addtolength{\baselineskip}{1.5mm} \thispagestyle{empty}

\vspace{48pt}

\setcounter{footnote}{0}

\newpage

\section{Introduction}

The quantum theory of gauge invariant cosmological perturbations is based on
the validity of general relativity and quantum field theory. Both of these
theories break down at Planckian scales. However if inflation lasts a little
bit longer than what is required to solve the problems of standard
cosmology- as predicted by most inflationary models \cite{Linde}- many
scales of cosmological size today have been sub-planckian at the onset of
inflation. So it is natural to ask if the present cosmic microwave spectrum
carries any thumbprint of physics at such small scales.

A number of papers have investigated the robustness of predictions of
inflation to transplanckian physics by introducing non-linearities into the
dispersion relation of Fourier modes \cite{Martin,
Niemeyer,Kowalski,Parentani}. More general arguments \cite{Kempf1}, which
may apply to any theory of quantum gravity \cite{Kempf2}, suggest a scenario
in which the ultraviolet cutoff is modelled by a modified quantum mechanical
commutation relation \cite{KMM}:
\begin{equation}
\lbrack \mathbf{X},\mathbf{P}]=i\hbar (\mathbf{1}+\beta \mathbf{P}^{2})
\label{1}
\end{equation}%
This uncertainty relation has appeared in various studies of string theory %
\cite{Veneziano}. Easther \textit{et.al.} \cite{Easther1,Easther2} solved
the equation for tensor perturbations numerically and found out that the
effect on the tensor power spectrum is of order $\sigma =\sqrt{\beta }H$,
where $\sqrt{\beta }$ is the minimal length associated with the hypothesized
ultraviolet cutoff and $H$ is the Hubble constant during inflation.

However it was recently shown that this approach has an ambiguity: the
presence of a cutoff not only affects the bulk terms of the Lagrangian
density but also the boundary terms \cite{Amjad1}. Hence a total time
derivative added to the classical action will not remain a total time
derivative in the presence of a cutoff. In general, it will lead to a
modification of the equations of motion. In a recent paper we exploited the
aforementioned modified commutation relation and the ambiguity associated
with it to explain the origin of cosmic-scale primordial magnetic fields %
\cite{Amjad2}.

Vacuum fluctuations of the inflaton, $\phi _{0}$ -- the field that
drives inflation -- produce both scalar and tensor perturbations,
both of which contribute to the anisotropy of the cosmic microwave
background radiation. For any inflationary model one can calculate
$r$, the ratio of tensor to scalar amplitudes. $r$ multiplies the
upper bound on the scalar density perturbations by a factor of
$(1+r)^{-1/2}$. By knowing it one can tighten the bounds on the
scalar spectral index \cite{Liddle,Salopek}. It is therefore
important to know $r$ in as much detail as possible in order to
extract cosmological parameters with more precision.

The effect of transplanckian physics on the tensor/scalar ratio
was addressed for the first time in \cite{Hui}, where the authors
discovered that the ratio will be influenced by the short distance
physics, if tranplanckian physics does not lead to the same vacuum
for scalar and tensor fluctuations. In this article, following the
discovery of \cite{Amjad1}, we explore how the non-minimal choices
of the boundary term for tensor and scalar fluctuations affect the
tensor/scalar ratio. The structure of our paper is as follows:
first, we present the equations that scalar and tensor
fluctuations satisfy in the presence of a UV cutoff \cite{Amjad1},
categorizing various cases for which the ratio can change.
Following ref. \cite{Easther1} we then solve these equations for
scalar and tensor perturbations numerically in a near de-Sitter
background. We compute how the scalar power spectrum varies as a
function of $\sigma$. In the fourth section we ultimately find the
ratio of tensor to scalar fluctuations.

\section{Ratio of Tensor/Scalar Fluctuations with a Cutoff}

Consider the action
\begin{equation}
S=\frac{1}{2}\int (\partial _{\mu }\phi \partial ^{\mu }\phi -V(\phi ))\sqrt{%
-g}~d^{4}x-\frac{1}{16\pi G}\int R\sqrt{-g}~d^{4}x  \label{2}
\end{equation}%
which describes a scalar inflaton field minimally coupled to gravity. \ We
assume that the metric $ds^{2}=a^{2}(\tau )\left( {d\tau }^{2}-{\delta }%
_{ij}d{y}^{i}d{y}^{j}\right) $ describes the background, which is an
homogenous isotropic Friedmann universe with zero spatial curvature.

We can decompose perturbations of the metric tensor into scalar, vector and
tensor modes in the usual way according to their transformation properties
under spatial coordinate transformations on the constant-time hypersurfaces.
We shall be concerned with the scalar and tensor modes here, whose
perturbative decomposition is given by
\begin{eqnarray}
ds_{S}^{2} &=&a^{2}(\tau )\left( (1+2\Phi )d\tau ^{2}-2\partial
_{i}Bdy^{i}d\tau -[(1-2\Psi )\delta _{ij}+2\partial _{i}\partial
_{j}E]dy^{i}dy^{j}\right)  \label{scalar} \\
ds_{T}^{2} &=&a^{2}(\tau )\left( d\tau ^{2}-[\delta
_{ij}+h_{ij}]dx^{i}dx^{j}\right)  \label{tensor}
\end{eqnarray}%
where $\Phi ,B,\Psi $ and $E$ are scalar fields and $h_{ij}$ is a symmetric
three-tensor field satisfying $h_{i}^{i}=0=h_{ij}{}^{,j}$. Fluctuations of
the inflation field are given by $\phi (y,\tau )=\phi _{0}(\tau )+\delta
\phi (y,\tau )$, where $\phi _{0}(\tau )$ is the homogenous part that drives
the background expansion, with $|\delta \phi |\ll \phi _{0}$.

Using eqs. (\ref{scalar},\ref{tensor}) and writing
\begin{equation}
\Re =-\frac{a^{\prime }}{a}\frac{\delta \phi }{\phi _{0}^{\prime }}-\Psi ,
\end{equation}%
which is the gauge-invariant intrinsic curvature, the action can be written
as \cite{Amjad1}
\begin{equation}
S_{S}^{(1)}=\frac{1}{2}\int d\tau ~d^{3}\mathbf{y}~z^{2}\left( (\partial
_{\tau }\Re )^{2}-\delta ^{ij}~\partial _{i}\Re \partial _{j}\Re \right) .
\label{scalar2}
\end{equation}%
or alternatively as
\begin{equation}
S_{S}^{(2)}=\frac{1}{2}\int d\tau d^{3}\mathbf{y}\left( {\left( \partial
_{\tau }u\right) }^{2}-\delta ^{ij}~{\partial }_{i}u~{\partial }_{j}u+\frac{%
z^{\prime \prime }}{z}u^{2}\right) .  \label{scalar1}
\end{equation}%
where $\Re =-u/z$ \cite{Mukhanov},
\begin{equation}
z=\frac{a\phi _{0}^{\prime }}{\alpha }  \label{z}
\end{equation}

These two actions for scalar fluctuations are equivalent up to a boundary
term in absence of minimal length, with $S_{S}^{(2)}$ more commonly used in
the literature because of its similarity with the action of a massive free
scalar field in Minkowskian space-time. However the effective mass, $%
z^{\prime \prime }/z$ is time dependent.

When the generalized uncertainty principal (\ref{1}) is employed, $%
S_{S}^{(1)}$ and $S_{S}^{(2)}$ are no longer equivalent. Instead, they
respectively yield the following equations of motion for the Fourier
components of $u$, \cite{Amjad1}:
\begin{equation}
{u}_{\tilde{k}}^{\prime \prime }+\frac{{\kappa }^{\prime }}{{\kappa }}{u}_{%
\tilde{k}}^{\prime }+\left( \mu -\frac{z^{\prime \prime }}{z}-\frac{%
z^{\prime }}{z}\frac{{\kappa }^{\prime }}{{\kappa }}\right) {u}_{\tilde{k}%
}=0,  \label{scr1}
\end{equation}%
\begin{equation}
{u}_{\tilde{k}}^{\prime \prime }+\frac{{\kappa }^{\prime }}{{\kappa }}{u}_{%
\tilde{k}}^{\prime }+\left( \mu -\frac{z^{\prime \prime }}{z}\right) {u}_{%
\tilde{k}}=0,  \label{scr2}
\end{equation}%
where
\begin{eqnarray}  \label{mu}
\mu (\tau ,\rho ) &=&\frac{a^{2}\rho ^{2}}{(1-\beta \rho ^{2})^{2}}
\label{35} \\
\kappa (\rho ) &=&\frac{e^{3\beta \rho ^{2}/2}}{1-\beta \rho ^{2}}.
\end{eqnarray}%
Here $\rho $ is a parameter that plays the role of inverse wavelength \cite%
{Kempf3}. The difference in the above equations of motion is attributed to
the non-triviality of the manner in which minimal length affects the boundary terms.
The Scalar fluctuation amplitude is then defined as \cite{Lidsey}
\begin{equation}
A_{S}(k)\equiv \frac{2}{5}P_{S}^{1/2}=\frac{2}{5}\sqrt{\frac{k^{3}}{2\pi ^{2}%
}}\left| \frac{u_{\tilde{k}}}{z}\right| _{\tilde{k}/aH\rightarrow 0}.
\label{scrpower}
\end{equation}

Similarly, for tensor perturbations, among an infinite number of actions
that are equivalent in the absence of minimal length we \textit{choose} \cite%
{Amjad1} to start from
\begin{equation}
S_{T}^{(1)}=\frac{m_{Pl}^{2}}{64\pi }\int d\tau
d^{3}\mathbf{y}~a^{2}(\tau )~\partial _{\mu }h^{i}{}_{j}~\partial
^{\mu }h_{i}{}^{j}  \label{tensor1}
\end{equation}%
and
\begin{equation}
S_{T}^{(2)}=\frac{1}{2}\int d\tau d^{3}\mathbf{y}\left( \partial _{\tau
}P_{i}{}^{j}\partial ^{\tau }P^{i}{}_{j}-\delta ^{rs}{\partial }%
_{r}P_{i}{}^{j}{\partial }_{s}P^{i}{}_{j}+\frac{a^{\prime \prime }}{a}%
P_{i}{}^{j}P^{i}{}_{j}\right) ,  \label{tensor2}
\end{equation}%
that again differ from each other by a boundary term. As we will
see, one of these actions has a  minimal effect on tensor/scalar
ratio whereas the other one has a maximal effect. Also,
$S_{T}^{(1)}$ is the action one obtains by directly expanding the
Einstein-Hilbert action. Here
\begin{equation}
P^{i}{}_{j}(y)=\sqrt{\frac{m_{Pl}^{2}}{32\pi }}~a(\tau
)h^{i}{}_{j}(y) \label{P}
\end{equation}%
and $h_{ij}$ is the transverse traceless part of tensor perturbations of the
metric (\ref{tensor}).

The $\tilde{k}$-Fourier component of $P_{ij}$ (denoted $p_{\tilde{k}}$),
satisfies the following equation of motion using the cutoff modified $%
S_{T}^{(1)}$%
\begin{equation*}
{p}_{\tilde{k}}^{\prime \prime }+\frac{{\kappa }^{\prime }}{{\kappa }}{p}_{%
\tilde{k}}^{\prime }+\left( \mu -\frac{a^{\prime \prime }}{a}-\frac{%
a^{\prime }}{a}\frac{{\kappa }^{\prime }}{{\kappa }}\right) {p}_{\tilde{k}%
}=0,
\end{equation*}%
whereas it satisfies
\begin{equation}
{p}_{\tilde{k}}^{\prime \prime }+\frac{{\kappa }^{\prime }}{{\kappa }}{p}_{%
\tilde{k}}^{\prime }+\left( \mu -\frac{a^{\prime \prime }}{a}\right) {p}_{%
\tilde{k}}=0.  \label{tsr2}
\end{equation}%
if we employ the variational principal on the cutoff modified $S_{T}^{(2)}$.

We define the tensor amplitude as \cite{Lidsey}
\begin{equation}
A_{T}(k)\equiv \frac{1}{10}P_{T}^{1/2}=\frac{1}{10}\sqrt{\frac{k^{3}}{2\pi
^{2}}}\left| h_{\tilde{k}}\right| _{\tilde{k}/aH\rightarrow 0}
\label{tsrpower}
\end{equation}%
The ratio of tensor to scalar fluctuations and scalar spectral index are
respectively given by%
\begin{equation}
r\equiv \frac{A_{T}^{2}}{A_{S}^{2}},  \label{r}
\end{equation}%
\begin{equation}
n(k)-1 \equiv \frac{d\ln A_{S}^{2}(k)}{d\ln k}.  \label{n}
\end{equation}%
The effect of $r$ is to multiply the upper bound on the density
perturbations by a factor of $(1+r)^{-1/2}$ which in turn affects our
estimation of the scalar spectral index \cite{Salopek,Liddle}.

In the absence of minimal length, one can expand the ratio of
tensor/scalar fluctuations in terms of the slow roll parameters.
To first order it is \cite{Liddle,Lidsey,Stewart}%
\begin{equation}
\frac{A_{T}^{2}}{A_{S}^{2}}=\epsilon  \label{r-slowroll}
\end{equation}%
where
\begin{equation}
\epsilon \equiv \frac{3\phi _{0}^{2}}{2} {\left( V(\phi _{0})+\frac{1}{2}{\dot{%
\phi}_{0}}^{2}\right)}^{-1} =\frac{m_{Pl}^{2}}{4\pi }\left( \frac{H_{\phi }}{H}%
\right) ^{2}.  \label{eps}
\end{equation}%
is the first slow-roll parameter \cite{Lidsey}. Here, $\phi$
subscript denotes differentiation with respect to $\phi $. In the
presence of minimal length the relation (\ref{r-slowroll}), takes
the following form
\begin{equation}
\frac{A_{T}^{2}}{A_{S}^{2}}=\epsilon {\left| \frac{p_{k}}{u_{k}}\right| }%
_{k/aH\rightarrow 0}^{2}  \label{r-slowroll-gen}
\end{equation}

The ambiguity in choosing the actions for scalar and tensor fluctuations in
the presence of the minimum length is a new source of transplanckian effects
that can modify the tensor/scalar ratio $r$. In general we have four
possibilities:

\begin{description}
\item[I,II] If we choose either $(S_{S}^{(1)},S_{T}^{(1)})$ or $%
(S_{S}^{(2)},S_{T}^{(2)})$ as the actions describing scalar and tensor
fluctuations, the scalar modes $u_{\tilde{k}}$, and tensor modes $p_{\tilde{k%
}}$, satisfy differential equations that are as similar as possible. In
particular this implies that in the special cases of near-de-Sitter and
power-law inflation where $z^{\prime \prime }/z=a^{\prime \prime }/a$ (see %
\cite{Lidsey}) and $z^{\prime }/z=a^{\prime }/a$ (see Appendix), the
equations governing both scalar and tensor perturbations are identical.
Since metric and inflaton perturbations cannot be fully distinguished in a
gauge invariant manner, scalar and tensor modes should also obey the same
initial conditions, yielding $\left| \frac{p_{k}}{u_{k}}\right| =1$. The
distinction between cases I and II becomes apparent when the inflating
background deviates from the power-law and near-de-sitter backgrounds.

\item[III,IV] The other extreme is to select either of the pairs $
(S_{S}^{(1)},S_{T}^{(2)})$ or $(S_{S}^{(2)},S_{T}^{(1)})$ to describe the
situation. In these cases the modes $u_{\tilde{k}}$ and $p_{\tilde{k}}$,
satisfy differential equations of differing form even in near-de-Sitter and
power-law backgrounds. In particular the tensor amplitude is not just $%
\epsilon $ times the scalar amplitude in power-law backgrounds. In the next
section we present a complete analysis of the scalar and tensor spectra in
near-de-Sitter space. We will investigate how the ratio of tensor to scalar
perturbations varies as a function of $\sigma $, the ratio of minimal length
to Hubble length during inflation.
\end{description}

\section{Scalar perturbations with minimum length in near-de-Sitter space}

Curvature fluctuations arise because the value of the inflaton field is
coupled to the energy density of the vacuum energy driving inflation, i.e.
fluctuations in the inflaton field result in fluctuations in the expansion
rate at linear order in perturbation theory. This coupling is what creates
fluctuations in the intrinsic curvature scalar, which are then manifest as
density fluctuations. Since in de Sitter space fluctuations in the inflaton
field, $\phi $, are not coupled to fluctuations in the energy density, the
amplitude of density fluctuations is zero. A naive exploitation of the
formalism of refs. \cite{Mukhanov,Lidsey} implies that the expression for
density fluctuations is singular for de Sitter space. The reason that the
expression is singular is not because the density fluctuation amplitude is
singular, but because the foliation of space-time implicit in the choice of
gauge becomes singular.

Nevertheless, we can proceed in this manner by assuming that $\epsilon $ is
close to zero, i.e. that the background is arbitrarily close to the de
Sitter limit. Note that we are taking $H$, the Hubble parameter, to be very
small, since it is known from COBE that $P_{S}=H^{2}/\epsilon \simeq
const.\times 10^{-5}$.

We begin with an analysis of the scalar power spectrum, tracking the
normalized modes which are inside the horizon until they are far outside the
horizon, where their amplitude determines the perturbation spectrum. To this
end, we will solve the mode equation (\ref{scr1}) numerically. As in Ref.%
\cite{Easther1,Easther2}, we describe the initial evolution by an
approximate analytic solution, which we then evolve numerically to late
times.

In this section and in what follows, we first analyze the action $%
S_{S}^{(2)} $ and $S_{T}^{(1)}$ for scalar and tensor perturbations
respectively. In de Sitter space $a=-1/(H\tau )$ and $z^{\prime \prime
}/z=2/\tau ^{2}$. A mode with a fixed comoving wave number $\tilde{k}$
corresponds over time to increasing proper wave lengths. Each mode's proper
wave length corresponds to the Planck length at some time $\tau $ that
depends on $\tilde{k}$ and this is when the evolution of that mode begins.
This time is when $a^{2}(\tau _{\tilde{k}})\simeq \beta \tilde{k}^{2}$ and $%
\rho ^{2}=1/\beta $. At this initial time equation (\ref{scr1}) has an
irregular singular point.

Since de Sitter space is time-translation invariant, the equation can be
written in terms of the dimensionless parameter $w=\tilde{k}\tau $, in terms
of which all the modes evolve jointly:
\begin{equation}
\frac{d^{2}u_{\tilde{k}}}{dw^{2}}+n(w)\frac{du_{\tilde{k}}}{dw}+\Omega
^{2}(w)u_{\tilde{k}}(w)=0,  \label{65}
\end{equation}%
where
\begin{eqnarray}
n(w) &=&\frac{1}{\theta (\zeta (w))}\frac{d\theta (\zeta (w))}{dw}
\label{66} \\
\Omega ^{2}(w) &=&-\left( \frac{1}{\sigma ^{2}w^{2}}\frac{W(\zeta (w))}{%
(1+W(\zeta (w)))^{2}}+\frac{2}{w^{2}}\right)
\end{eqnarray}%
and in de Sitter space $\zeta (w)=-\sigma ^{2}w^{2}$. Here we define
\begin{equation}
\sigma =\sqrt{\beta }/H^{-1},
\end{equation}%
which is the ratio of the minimal length scale and the Hubble length scale
during inflation. The function $W(x)$ is the Lambert W-function, defined by
the relation $W(x)e^{W(x)}=x$ \cite{19}. \

Equation (\ref{65}) has a singular point at $w_{crit}=\varpi =-\frac{1}{%
\sigma \sqrt{e}}$. The singular point at $\varpi $ is an irregular singular
point because the coefficients of $du_{\tilde{k}}/dw$ and $u_{\tilde{k}}$
are not analytic in $v=w-\varpi $
\begin{eqnarray}
n(v) &=&-\frac{1}{2v}-\frac{7}{12}\frac{e^{1/2}}{\sqrt{Av}}+\frac{67}{144}%
\frac{e}{A}+\cdots  \notag  \label{67} \\
\Omega ^{2}(v) &=&\frac{A}{v}-\frac{e^{1/2}\sqrt{A}}{\sqrt{v}}-\frac{37}{72}%
e+\cdots
\end{eqnarray}%
where
\begin{equation}
A=\frac{e^{1/2}}{4\sigma }.  \label{68}
\end{equation}%
Proceeding along the lines given in ref. \cite{Easther1,Easther2}, we solve
for the leading behavior of $u_{\tilde{k}}$ by extracting the most singular
terms of the equation of motion%
\begin{equation}
\ddot{u}_{\tilde{k}}-\frac{1}{2v}\dot{u}_{\tilde{k}}+\frac{A}{v}u_{\tilde{k}%
}=0.  \label{69}
\end{equation}%
where in the overdot now denotes the derivative with respect to $v$.
Ignoring the $\dot{u}_{\tilde{k}}$ term, this equation is similar to the
high frequency limit of the mode equation:
\begin{equation}
u_{k}^{\prime \prime }(\tau )+\Omega _{k}^{2}(\tau )u_{k}(\tau )=0
\label{heq}
\end{equation}%
whose solution is approximated by the WKB form
\begin{equation}
u_{k}(\tau )=\frac{1}{\sqrt{2\Omega _{k}}}\exp (-i\int^{\tau }\Omega
_{k}(\tau ^{\prime })d\tau ^{\prime })  \label{70}
\end{equation}%
if the adiabatic condition $|\Omega _{k}^{\prime }/\Omega
_{k}^{2}|\ll 1$ is satisfied. This choice of vacuum, which is
called Bunch-Davies vacuum, reduces to the Minkowskian vacuum for
wavelengths smaller than the Hubble scale. Inspired by this
similarity, one can suggest a Bunch-Davies-like vacuum of the
form:
\begin{equation}
u_{\tilde{k}}(v)=(\frac{v}{4{\tilde{k}}^{2}A})^{1/4}\exp (-2i\sqrt{Av})
\label{71}
\end{equation}%
with
\begin{equation}
\Omega _{\tilde{k}}=\tilde{k}\sqrt{A/v}.  \label{72}
\end{equation}%
This vacuum does not satisfy the adiabaticity conditions in the vicinity of
its creation time, $v=0$. To be specific:
\begin{equation}
\frac{\Omega _{\tilde{k}}^{\prime }}{\Omega _{\tilde{k}}^{2}}=\frac{\tilde{k}%
}{2}\frac{1}{\sqrt{Av}}  \label{73}
\end{equation}%
For $v\sim 0$ the adiabatic condition is violated. It means that each mode
is born in an excited state. In the model of transplanckian physics proposed
in ref.\cite{Martin}, each mode undergoes three phases in its evolution. In
the first phase, the wavelength of the given mode is much smaller than the
Planck length: $\lambda \ll l_{p}$. Each mode is born into the vacuum state
that minimizes the Hamiltonian and satisfies the adiabaticity condition. In
the second phase, the wavelength of the mode is larger than the Planck
length but still smaller than the Hubble radius, $l_{p}\ll \lambda \ll l_{H}$%
. In the third phase the mode is outside the Hubble radius: $\lambda \gg
l_{H}$. In our version of this scenario, the first phase is removed and
replaced by an excited initial state which violates the adiabaticity
conditions.

In fact, equation (\ref{69}) is solved exactly by
\begin{equation}
u(y)=C_{+}F(v)+C_{-}F^{\ast }(v)  \label{74}
\end{equation}%
where
\begin{equation}
F(y)=(\frac{\sqrt{A}}{2}+iA\sqrt{v})\exp (-2i\sqrt{Av})  \label{75}
\end{equation}%
and where the coefficients $C_{\pm }$ are constrained through the Wronskian
condition:
\begin{equation}  \label{Wronskian2}
u_{\tilde{k}}(\tau )u_{\tilde{k}}^{\ast ^{\prime }}(\tau )-u_{\tilde{k}%
}^{\ast }(\tau )u_{\tilde{k}}^{\prime }(\tau )=i{\kappa} ^{-1},
\end{equation}
that reduces to
\begin{equation}
|C_{+}|^{2}-|C_{-}|^{2}=\frac{e^{-1}}{2\tilde{k}A^{3}}.  \label{76}
\end{equation}%
This equation will lead to a one parameter family of solutions.
Similarity with the Bunch-Davies vacuum suggests that $C_{-}=0$.
In addition, in this case, it is possible to obtain conventional
QFT result when $\sigma \rightarrow 0$. However there exist other
legitimate choices of the vacuum. Specifically, it is possible to
recover the standard QFT result in the limit
 $\sigma \rightarrow 0$, if $C_{-}$ approaches zero faster than $\sigma
^{3/2}$, as we shall subsequently demonstrate. However we will first assume
that $C_{-}=0$.
\begin{figure}[t]
\includegraphics[angle=270, scale=0.35]{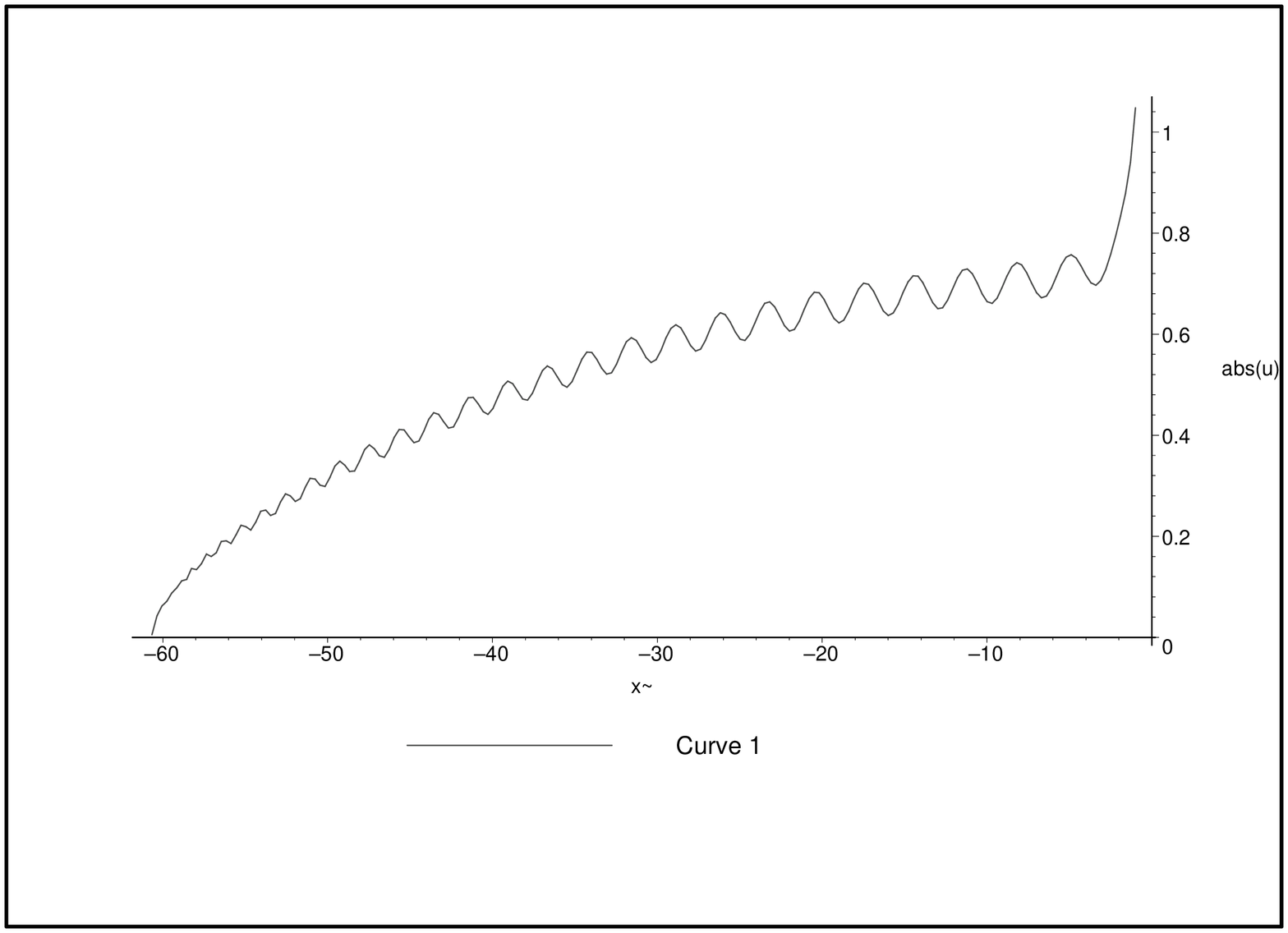} \notag %
\includegraphics[angle=270, scale=0.35]{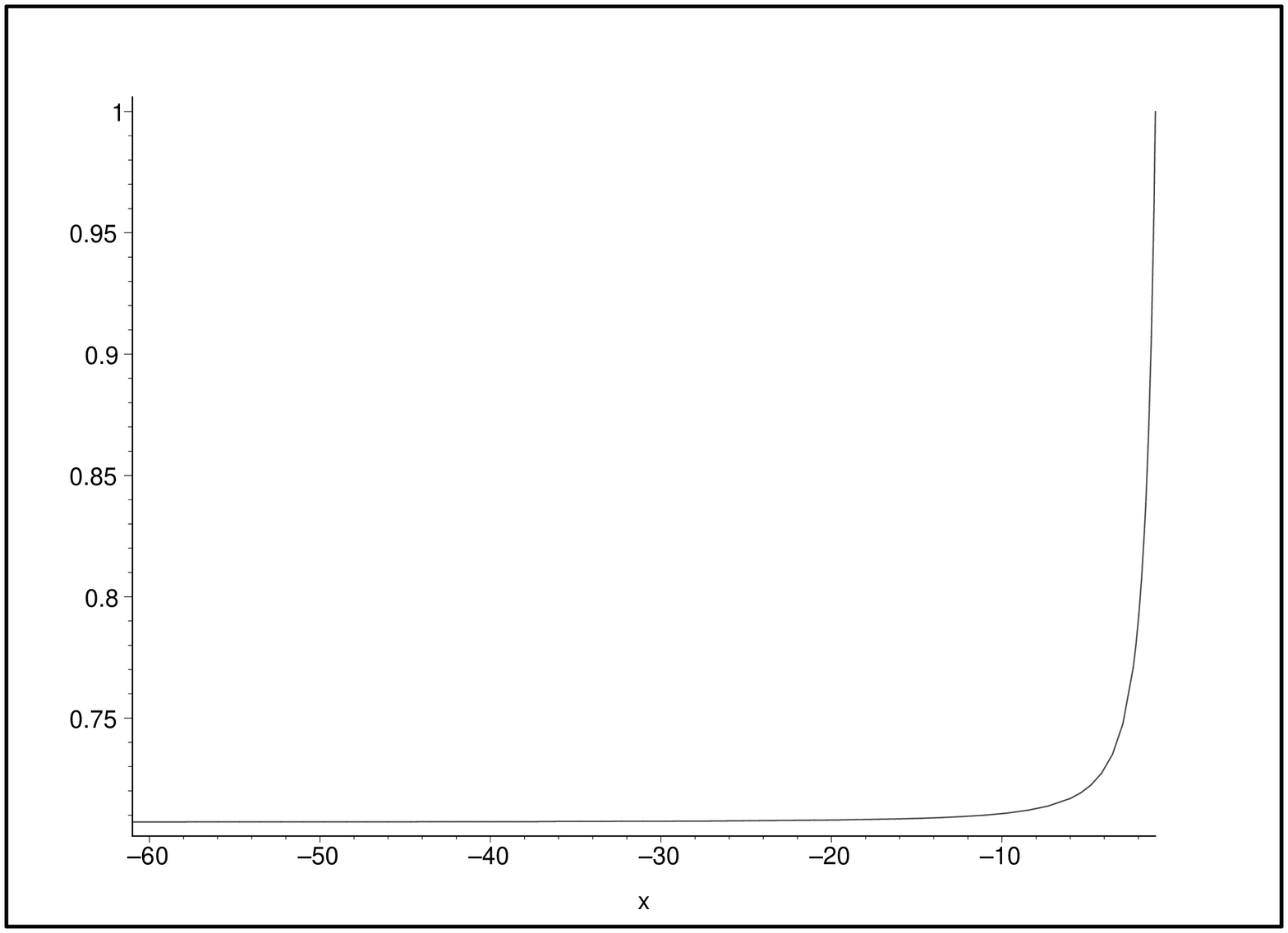}
\caption{The left figure shows how each mode evolves in the presence of a
cutoff as a function of $w=\tilde{k}\protect\tau $. Each mode is created at
a finite conformal time and the amplitude of the modes is modulated on a
monotonously increasing curve. $C_{-}$ is assumed to be zero. The right
figure shows how in absence of minimal length each mode is created at
infinite conformal time and its amplitude monotonously increases until it
leaves the horizon.}
\label{fig1}
\end{figure}

Equation (\ref{65}) has been solved to order $1/v$. As in Ref.\cite{Easther1}
we will extract the subleading behavior of $u_{\tilde{k}}$ by the method of
dominant balance \cite{22}. We solve the differential equation (\ref{65}) up
to $1/\sqrt{v}$ by defining $u_{\tilde{k}}(v)=F(v)(1+\epsilon _{1}(v))$,
extracting the most singular terms. The equation of motion for $\epsilon
_{1} $ is:
\begin{equation}
\frac{d^{2}\epsilon _{1}}{dv^{2}}-\frac{1}{2v}\frac{d\epsilon _{1}}{dv}-%
\frac{3}{2}e^{1/2}\sqrt{\frac{A}{v}}=0  \label{77}
\end{equation}%
which has the solution
\begin{equation}
\epsilon _{1}(v)=\frac{1}{3}\sqrt{A}e^{1/2}v^{3/2}(3\ln (v)-2).  \label{78}
\end{equation}%
The solution is improved by extracting the residual $\ln (v)$ terms. To do
so, we replace $u_{\tilde{k}}$ by $F(v)(1+\epsilon _{1}(v))(1+\epsilon
_{2}(v))$ and extract the most singular terms to obtain the following
differential equation for $\epsilon _{2}(v)$
\begin{equation}
\frac{d^{2}\epsilon _{2}}{dv^{2}}-\frac{1}{2v}\frac{d\epsilon _{2}}{dv}-%
\frac{7}{8}e\ln (v)=0.  \label{79}
\end{equation}%
whose solution is%
\begin{equation}
\epsilon _{2}(v)=\frac{7}{16}ev^{2}(2\ln (v)-5).  \label{80}
\end{equation}

We have solved the differential equation (\ref{65}) up to terms that vanish
as $v\rightarrow 0$. We glue this analytic solution, which is valid when the
mode is in the vicinity of the irregular singular point and inside the
horizon, to the full numerical evaluation of the mode equation. As the
coefficients of $u_{\tilde{k}}$ and $u_{\tilde{k}}^{\prime }$ are infinite
at $v=0$, this junction is done at a finite nonzero value of $v_{0}$. By
varying $v_{0}$ we have checked that our results do not depend on the choice
of starting point. We evolved the mode equation using Fehlberg fourth-fifth
order Runge-Kutta method implemented in Maple $9$. In Figure (1) we have
shown how each fluctuation mode evolves as a function of $w=\tilde{k}\tau $.
In the absence of minimal length there is no birth time for the modes and $%
|u_{k}|$ increases monotonically as it evolves. As we incorporate minimal
length into the problem, each mode is created at a definite $\tilde{k}$%
-dependent time and $|u_{\tilde{k}}|$ is modulated on a monotonically
increasing function until it crosses the horizon. At that time $|u_{k}|$
stops oscillating and goes to infinity as we approach the present time. One
should notice that parameter $\tilde{k}$ is different from the comoving
momentum at large momenta. This difference modifies the condition of horizon
crossing in terms of the parameter $\tilde{k}$. Note that $\rho $ plays the
role of inverse wavelength in our model \cite{Kempf3}. Using the relation
between $\rho $ and $\tilde{k}$ \cite{Amjad1},
\begin{equation}
\tilde{k}^{i}=a\rho ^{i}e^{-\beta \rho ^{2}/2},  \label{rho-k}
\end{equation}%
we can express the criterion of horizon crossing , $\rho =H$, in terms of
parameter $\tilde{k}$
\begin{figure}[t]
\includegraphics[angle=270, scale=0.35]{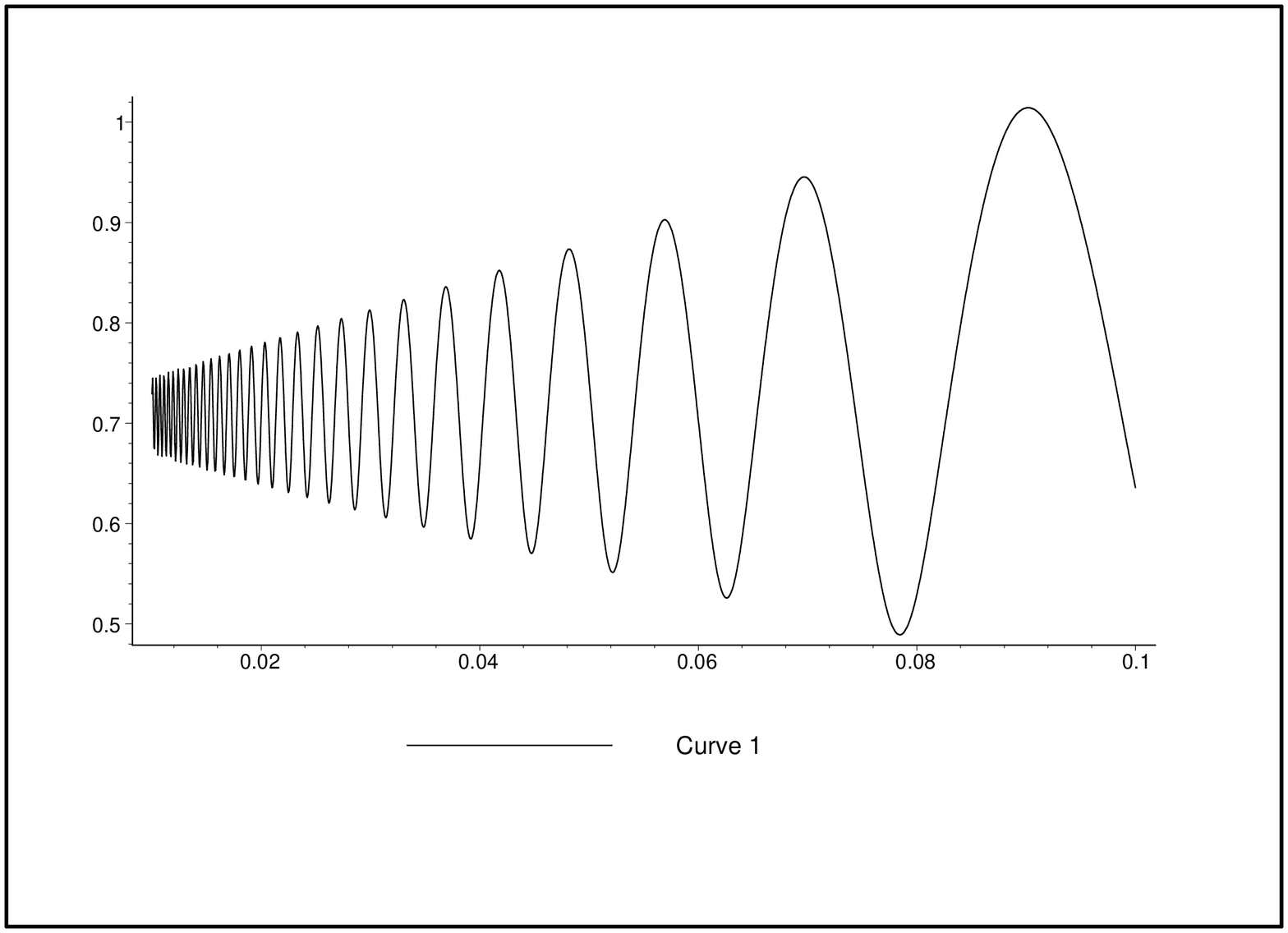} %
\includegraphics[angle=270, scale=0.35]{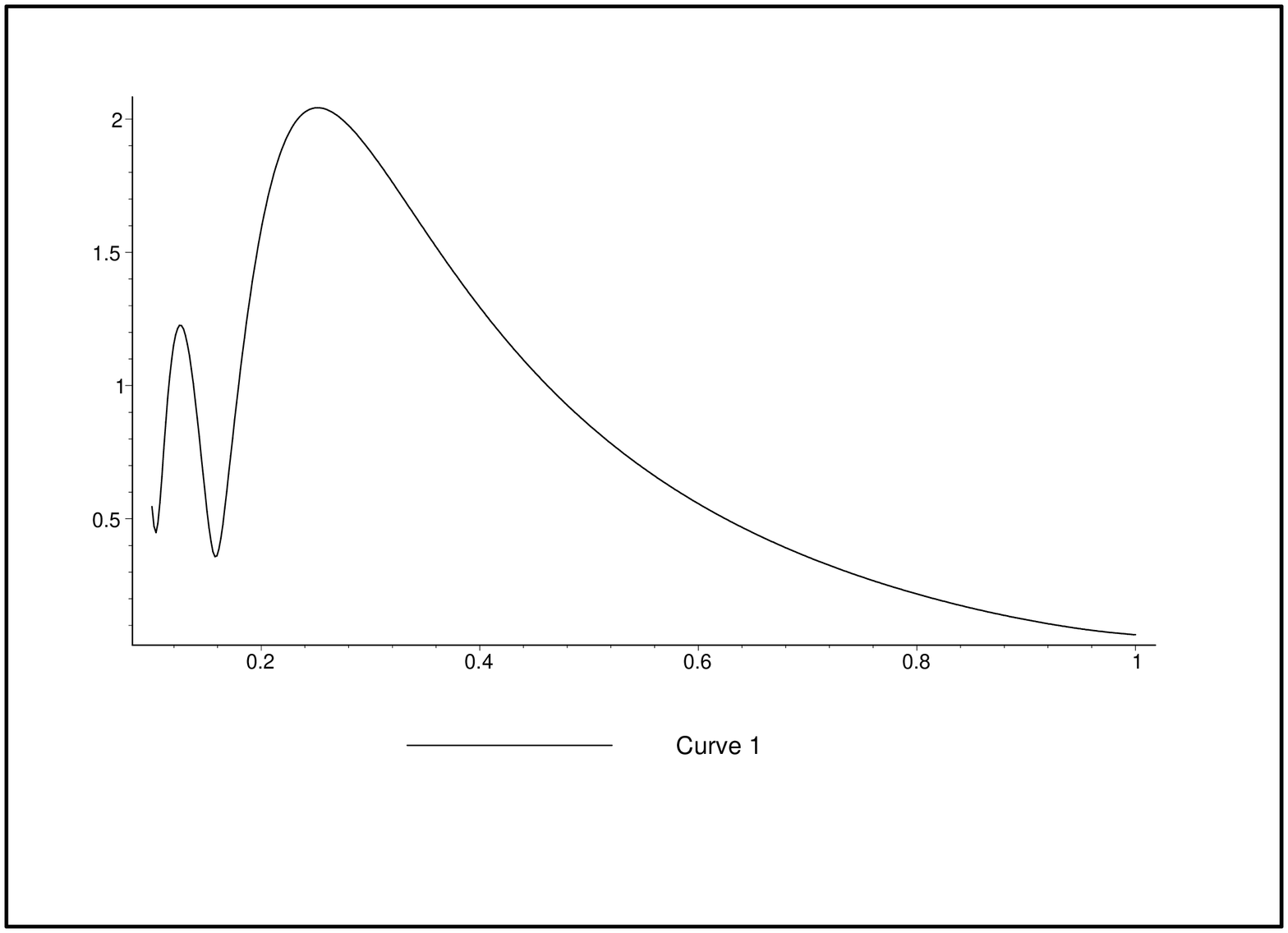}
\caption{The dependence of $\frac{\protect\sqrt{2}{\protect\pi }\dot{\protect%
\phi}}{H^{2}}P_{S}^{1/2}$ is plotted against $\protect\sigma $. As $\protect%
\sigma $ goes to zero, the standard result of $\frac{1}{\protect\sqrt{2}}$
is obtained. It is assumed that $C_{-}=0$.}
\label{fig2}
\end{figure}
\begin{equation}
\tilde{k}=aH\exp (-\beta H^{2}/2)  \label{81}
\end{equation}%
which takes the following form in de Sitter space
\begin{equation}
w_{\text{horizon}}=-\exp (-\sigma ^{2}/2)  \label{82}
\end{equation}%
where $w=\tilde{k}\tau $. In absence of minimal length this criterion
reduces to the familiar one in de Sitter space, $w=-1$.
\begin{figure}[t]
\includegraphics[angle=270, scale=0.35]{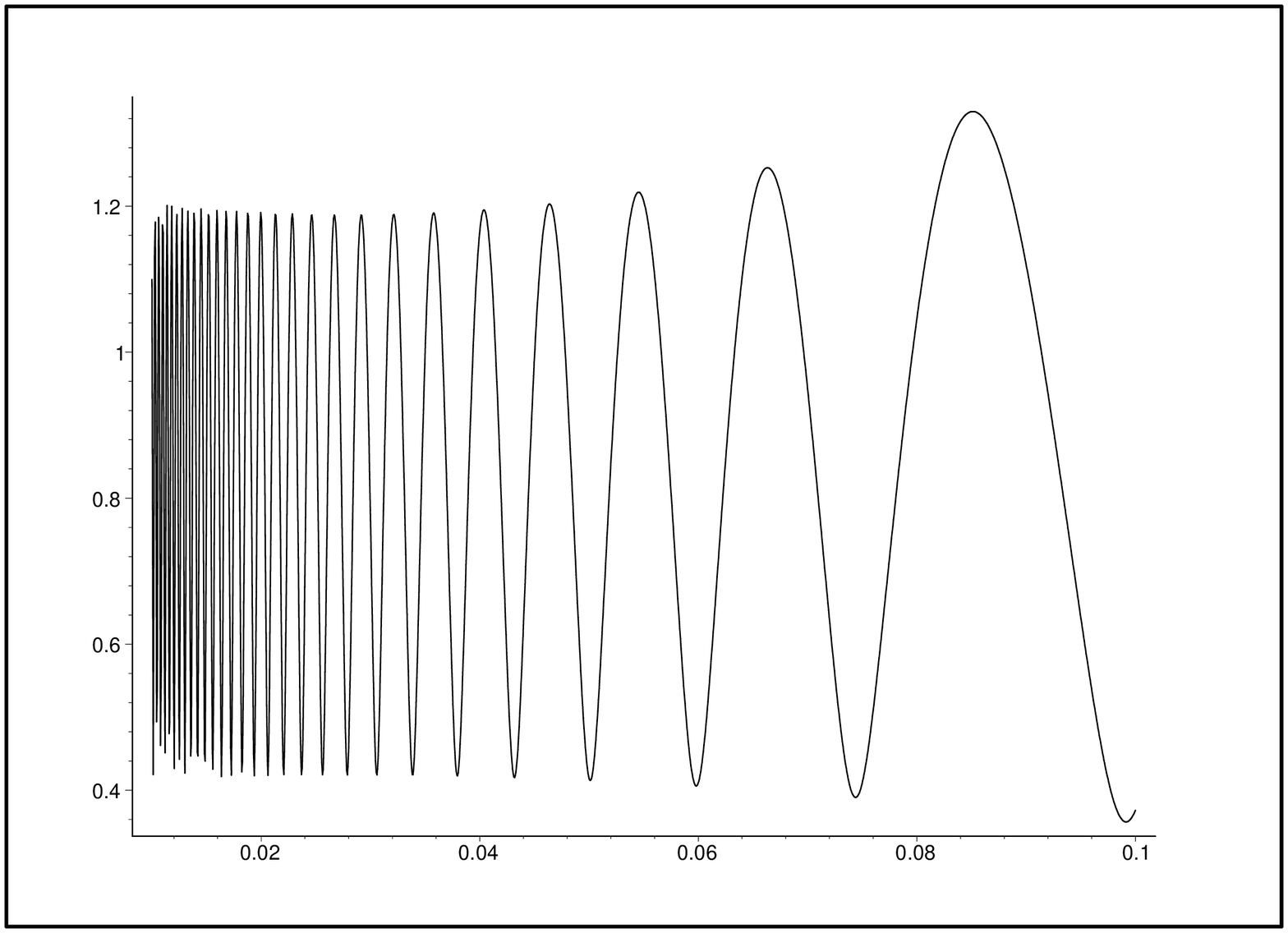} %
\includegraphics[angle=270, scale=0.35]{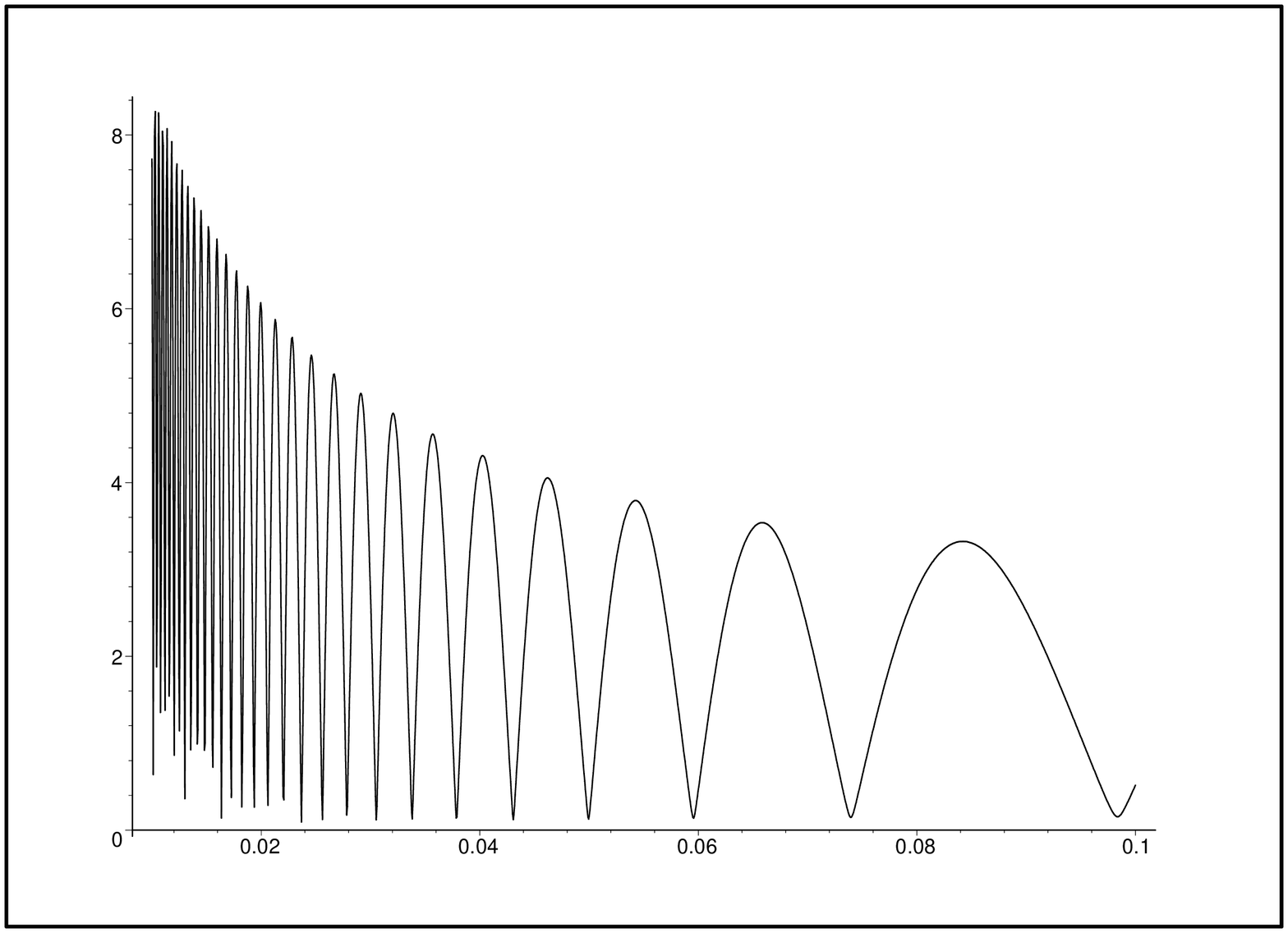}
\caption{The left figure shows the dependence of $\frac{\protect\sqrt{2}{%
\protect\pi }\dot{\protect\phi}}{H^{2}}P_{S}^{1/2}$ against $\protect\sigma $
when $C_{-}/C_{+}=0.5$. The right figure graphs the dependence of $\frac{%
\protect\sqrt{2}{\protect\pi }\dot{\protect\phi}}{H^{2}}P_{S}^{1/2}$ on $%
\protect\sigma $ when $C_{-}\propto \protect\sigma $. In neither of these
cases do we recover the standard field theory result when $\protect\sigma %
\rightarrow 0$.}
\label{fig3}
\end{figure}

For values of $\sigma $ close to $1$, i.e. when the energy scale of
inflation is of the order of the minimal length, the horizon-crossing
condition is considerably modified. Of course, we are really interested in
the asymptotic values of $|u_{\tilde{k}}|$, when $\rho /H\rightarrow 0$. To
implement it numerically, we have assumed this condition is satisfied when $%
\rho /H=0.01$. We can express this condition in terms of parameter $w$:
\begin{equation}
w_{\text{asymp}}=-0.01\exp (-\sigma ^{2}/20000).  \label{83}
\end{equation}%
The general answer to Equation(\ref{65}), has the form of $u_{\tilde{k}%
}(\tau )=N(\tilde{k})U_{\tilde{k}}(w)$. Comparison with Equation (\ref{76})
yields $N(\tilde{k})=1/\sqrt{\tilde{k}}$. So, the power spectrum in near-de
Sitter space is:
\begin{equation}
P_{S}^{1/2}=\sqrt{\frac{\tilde{k}^{3}}{2\pi ^{2}}}\left| \frac{u_{\tilde{k}}%
}{z}\right| =\frac{H^{2}}{\dot{\phi}}\frac{\left| -wU_{\tilde{k}}(w)\right|
}{\pi \sqrt{2}}|_{w=w_{\text{asymp}}}  \label{84}
\end{equation}%
On the other hand, quantum field theory yields the following result for the
near-de Sitter space
\begin{equation}
P_{S}^{1/2}(\sigma =0)=\frac{H^{2}}{2\pi \dot{\phi}}  \label{85}
\end{equation}%
\begin{figure}[t]
\includegraphics[angle=270, scale=0.35]{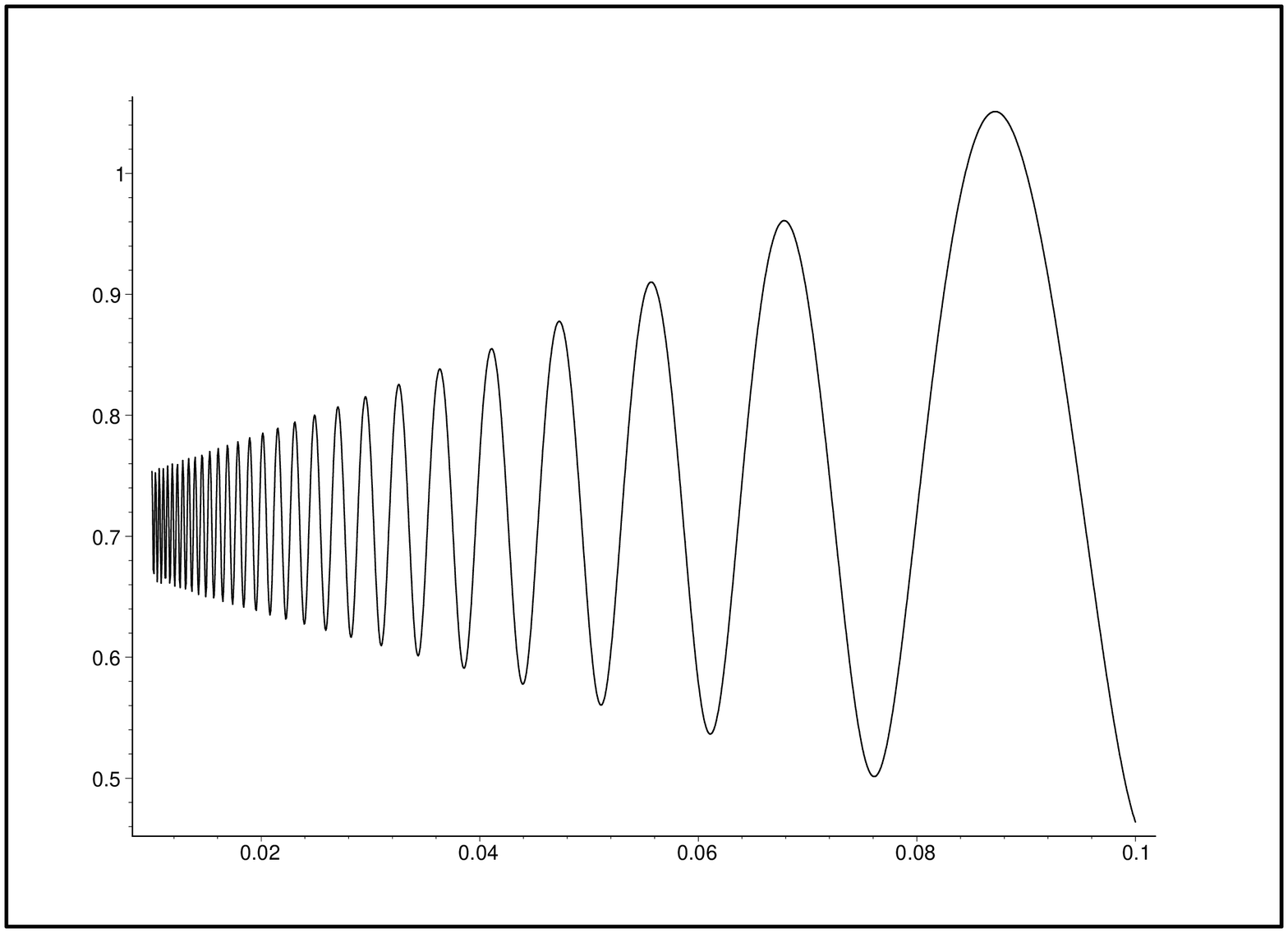} %
\includegraphics[angle=270, scale=0.35]{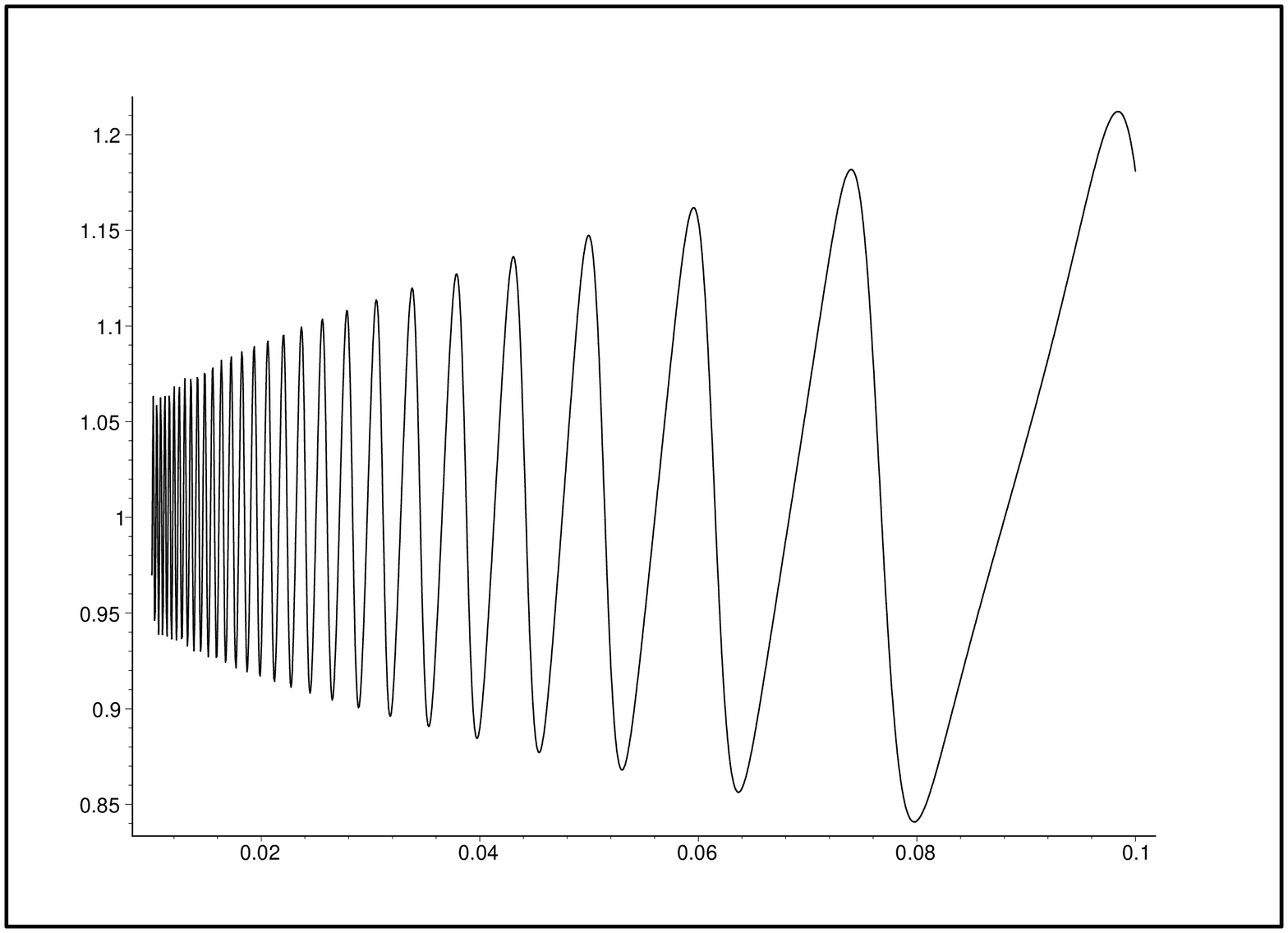}
\caption{The left figure shows the dependence of $\frac{\protect\sqrt{2}{%
\protect\pi }}{H}P_{S}^{1/2}$ against $\protect\sigma $ when $C_{-}\propto
\protect\sigma ^{2}$. The right figure shows the ratio of scalar power
spectrum when $C_{-}=0$ to scalar power spectrum when $C_{-}\propto \protect%
\sigma ^{2}$. In this case the scalar power spectrum approaches the standard
result when $\protect\sigma \rightarrow 0$}
\label{fig4}
\end{figure}

In Fig.$2$ we have displayed the $\sigma $ dependence of the scalar power
spectrum when $C_{-}=0$. For small values of $\sigma $, the power spectrum
has an oscillatory behavior around its standard result. For sufficiently
large values of $\sigma $ the power spectrum is significantly suppressed.
This could be used as a mechanism for solving the fine tuning problem of
inflationary models \cite{Easther1}.

Next we relax the condition $C_{-}=0$. Equation (\ref{76}) will lead to a
one parameter family of solutions. The criterion of approaching the result
of conventional quantum field theory when $\sigma \rightarrow 0$ can be used
to constrain our space of solutions. It has been pointed out \cite{Easther1}
that if the ratio of $C_{-}/C_{+}$ is a non-zero constant then the tensor
power spectrum does not approach its standard result in the limit $\sigma
\rightarrow 0$ . In Fig.3, we have examined this statement for scalar
perturbations and noticed that the same thing happens for scalar
perturbations too. However, in general, $C_{-}$ can be a function of $\sigma
$. Since we know that when $C_{-}=0$ and $C_{+}\propto \sigma ^{3/2}$, we
obtain the standard result in the limit of $\sigma \rightarrow 0$, we expect
that if $C_{-}$ approaches zero faster than $\sigma ^{3/2}$ when $\sigma
\rightarrow 0$ the criterion of recovering the standard QFT result is
satisfied. We have verified this statement for $C_{-}\propto \sigma $ and $%
C_{-}\propto \sigma ^{2}$ respectively as shown in Figures 3 and 4. Hence we
conjecture that it is possible that $C_{-}$ be proportional to $\sigma ^{n}$%
, $n>\frac{3}{2}$ whilst obtaining the standard QFT result in the limit $%
\sigma \rightarrow 0$.

In most inflationary models the expansion rate is slower than de Sitter
space; the ratio of minimum length to physical horizon is not constant and
decreases towards the end of inflation. Our study of de Sitter space
suggests that the amplitude of the longer modes will be affected more.
Whether it has observable effects depends on the energy scale of inflation.
We plan to return to this problem in future work \cite{Amjad4}.

\section{Ratio of tensor to scalar perturbations with minimum length in
near-de Sitter space}

As explained above, if the action of the tensor perturbations, in
absence of $\sigma $, is given in eq.(\ref{tensor2}) then the
tensor power spectrum will be $\epsilon $ times that of the scalar
perturbations. Otherwise, if the action is given by eq.
(\ref{tensor1}), its power spectrum will not be a simple multiple
of the scalar perturbations. Although a complete analysis of the
equation of motion derived from this action has been done once in
de Sitter space \cite{Easther1}, we recapitulate those
calculations in the present context to find the ratio of tensor to
scalar fluctuations. The outline of the calculations is completely
similar to what was done for scalar perturbations: we tailor the
solution that is valid in vicinity of the irregular singular point
to the numerically integrated solution. The exact analytic
solution in the neighborhood of the singular point is:
\begin{figure}[t]
\includegraphics[angle=270, scale=0.35]{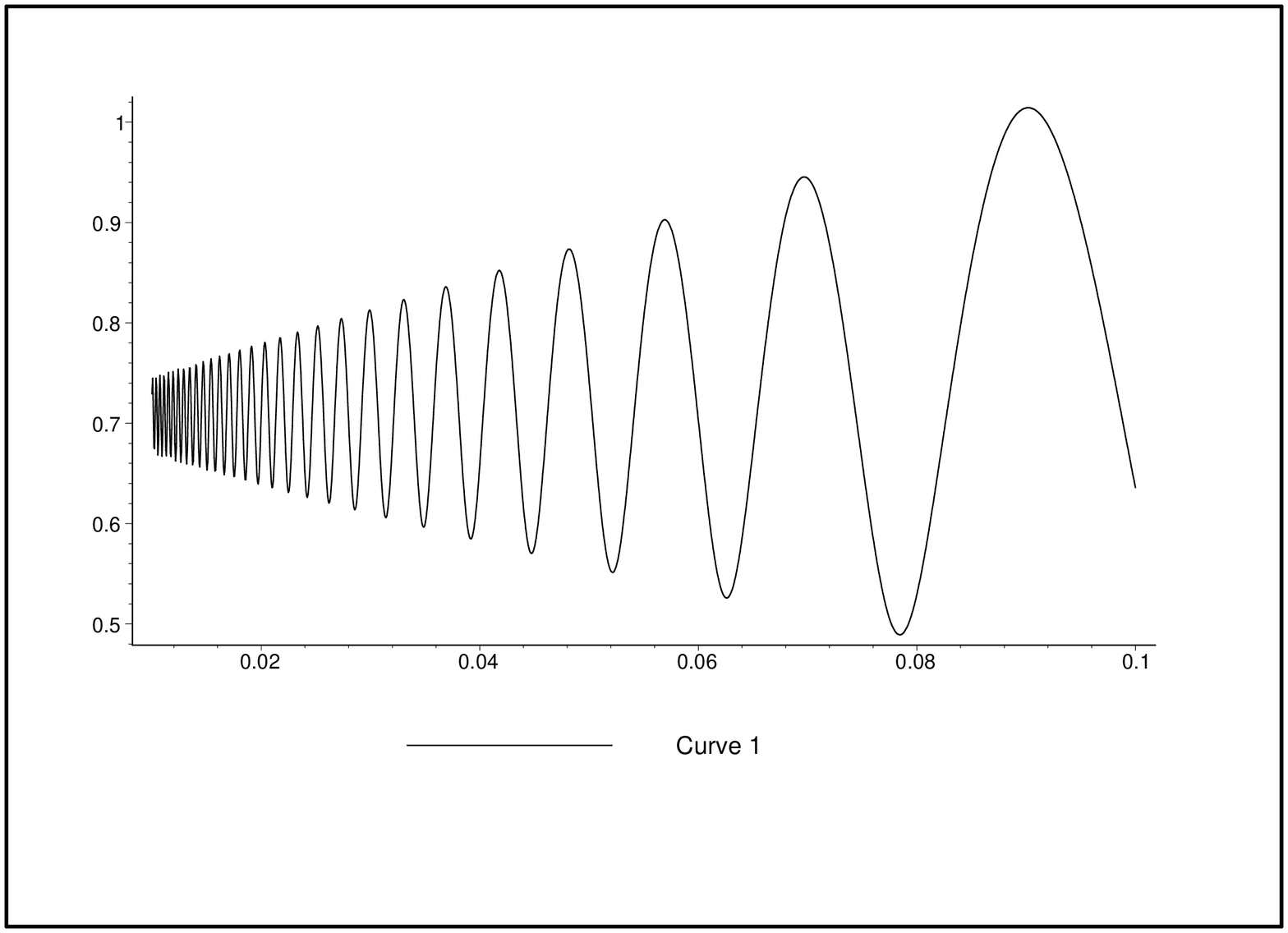} %
\includegraphics[angle=270, scale=0.35]{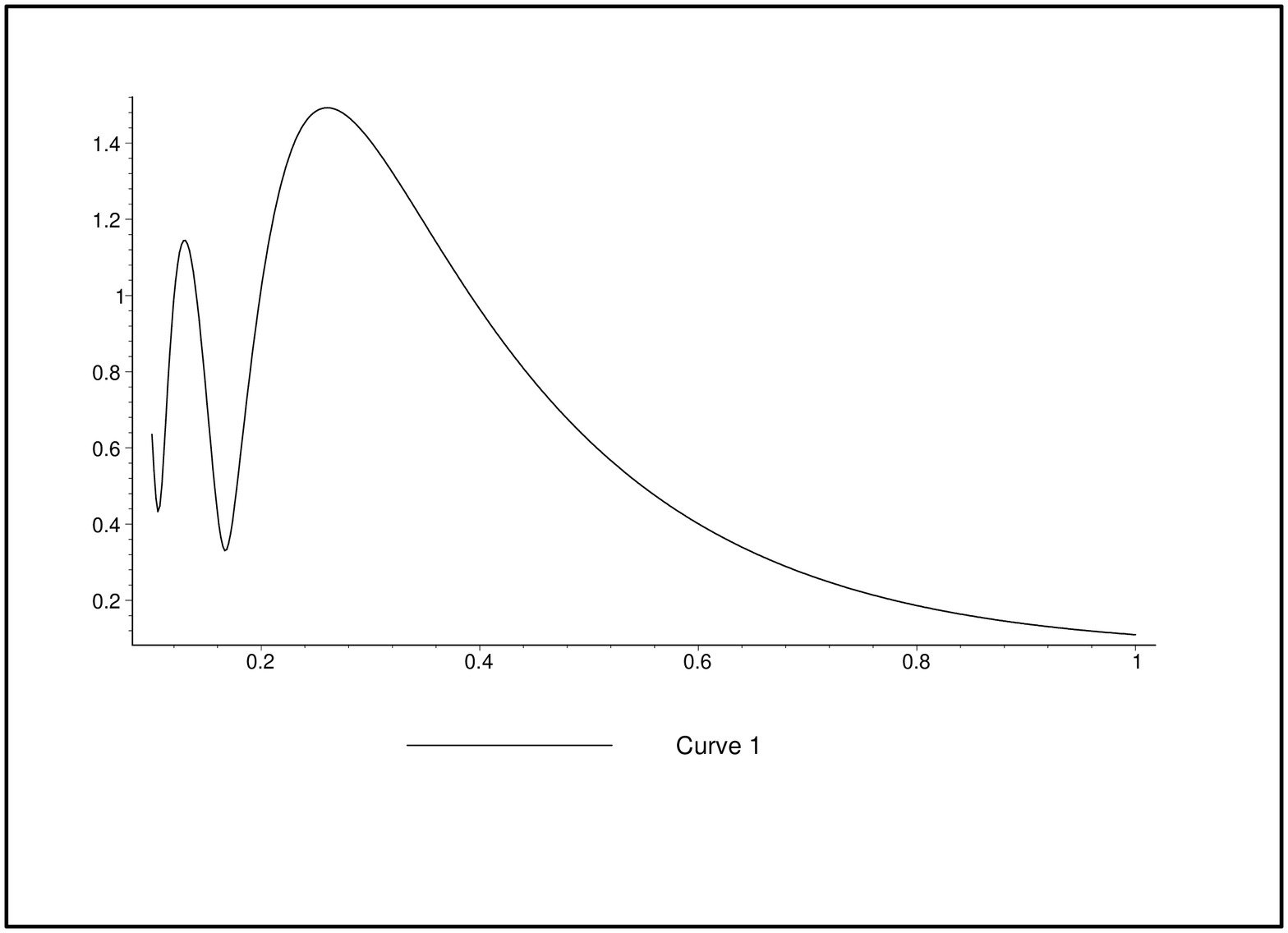}
\caption{The dependence of $\frac{\protect\sqrt{\protect\pi }m_{P}}{4H}%
P_{T}^{1/2}$ is plotted against $\frac{\protect\sqrt{\protect\pi }m_{P}}{4H}%
P_{T}^{1/2}$ approaches the standard result of $\frac{1}{\protect\sqrt{2}}$
as $\protect\sigma $ goes to zero. We have assumed that $D_{-}=0$.}
\label{fig5}
\end{figure}

\begin{equation}
u_{\tilde{k}}=D_{+}G(v)(1+\xi _{1}(v))(1+\xi _{2}(v))+D_{-}G^{\ast
}(v)(1+\xi _{1}(v))(1+\xi _{2}(v))  \label{86}
\end{equation}%
where
\begin{eqnarray}
G(y) &=&(\frac{\sqrt{B}}{2}+iB\sqrt{v})\exp (-2i\sqrt{Bv}),  \notag
\label{87} \\
\xi _{1}(v) &=&\frac{1}{3}e^{1/2}v^{3/2}\sqrt{A}(3\ln (v)-2),  \notag \\
\xi _{2}(v) &=&\frac{7}{8}ev^{2}\ln (v)-\frac{35}{16}ev^{2},
\end{eqnarray}%
\begin{equation}
B=\frac{1}{8}\frac{8A^{2}+e}{A}  \label{88}
\end{equation}%
$D_{+}$ and $D_{-}$ are constrained by the following wronskian condition
\begin{equation}
|D_{+}|^{2}-|D_{-}|^{2}=\frac{e^{-1}}{2\tilde{k}B^{3}}  \label{89}
\end{equation}%
Again, we have considerable freedom in choosing our vacuum. Since the
right-hand side of eq.(89) approaches zero like $\sigma ^{3}$ when $\sigma
\rightarrow 0$, if $D_{-}$ tends to zero faster than $\sigma ^{3/2}$ we can
recover the standard QFT result. Hereafter we restrict ourselves to $D_{-}=0$
so as to have a Bunch-Davies-like vacuum. We use equation (\ref{89}) with $%
D_{-}=0$ at a point close to the singularity to integrate the differential
equation. In Fig.5 we have displayed the $\sigma $-dependence of the tensor
power spectrum. The oscillationary behavior in vicinity of $\sigma =0$ and
decaying behavior for larger values of $\sigma $ has repeated.
\begin{figure}[t]
\includegraphics[angle=270, scale=0.35]{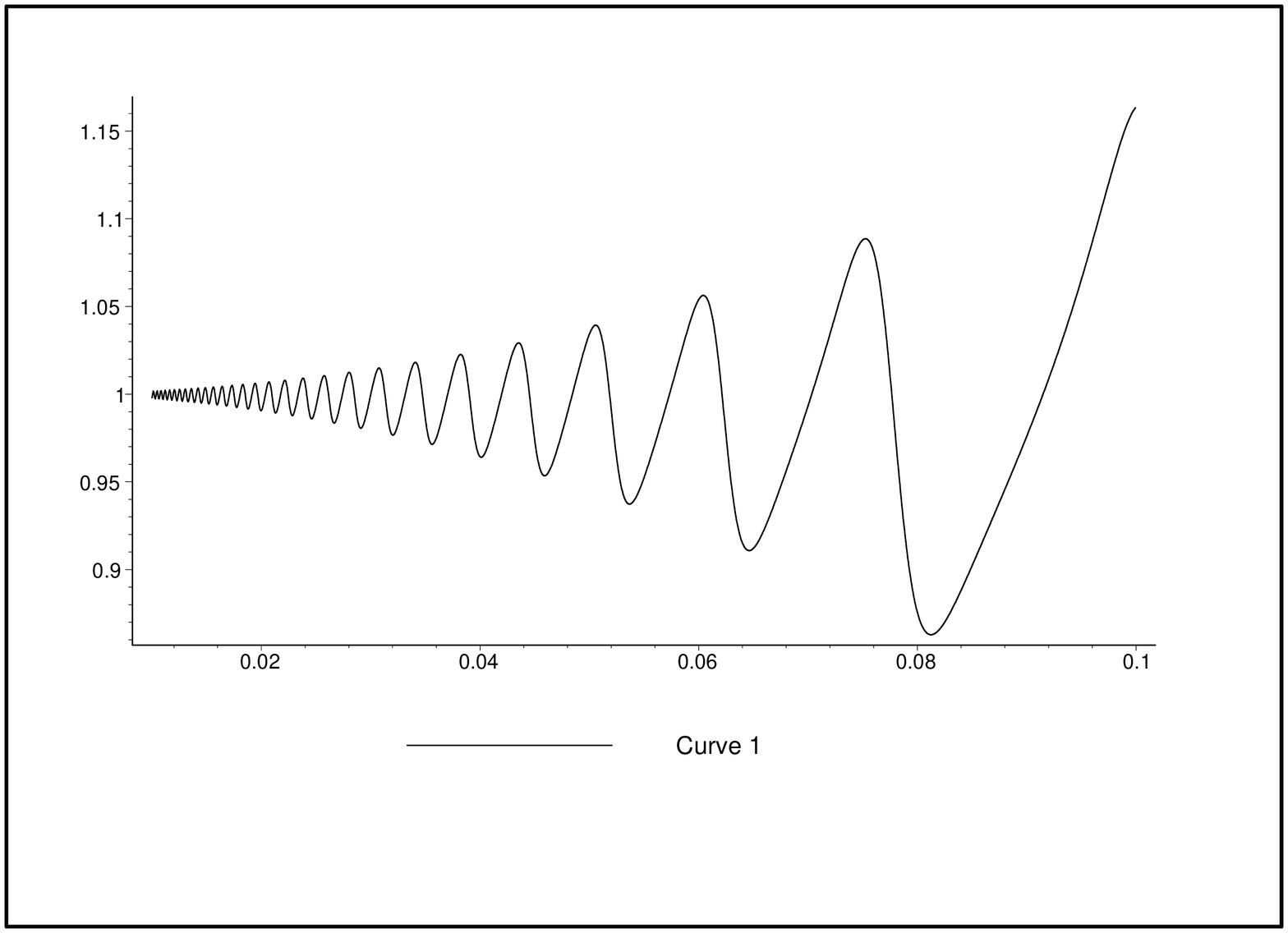} %
\includegraphics[angle=270, scale=0.35]{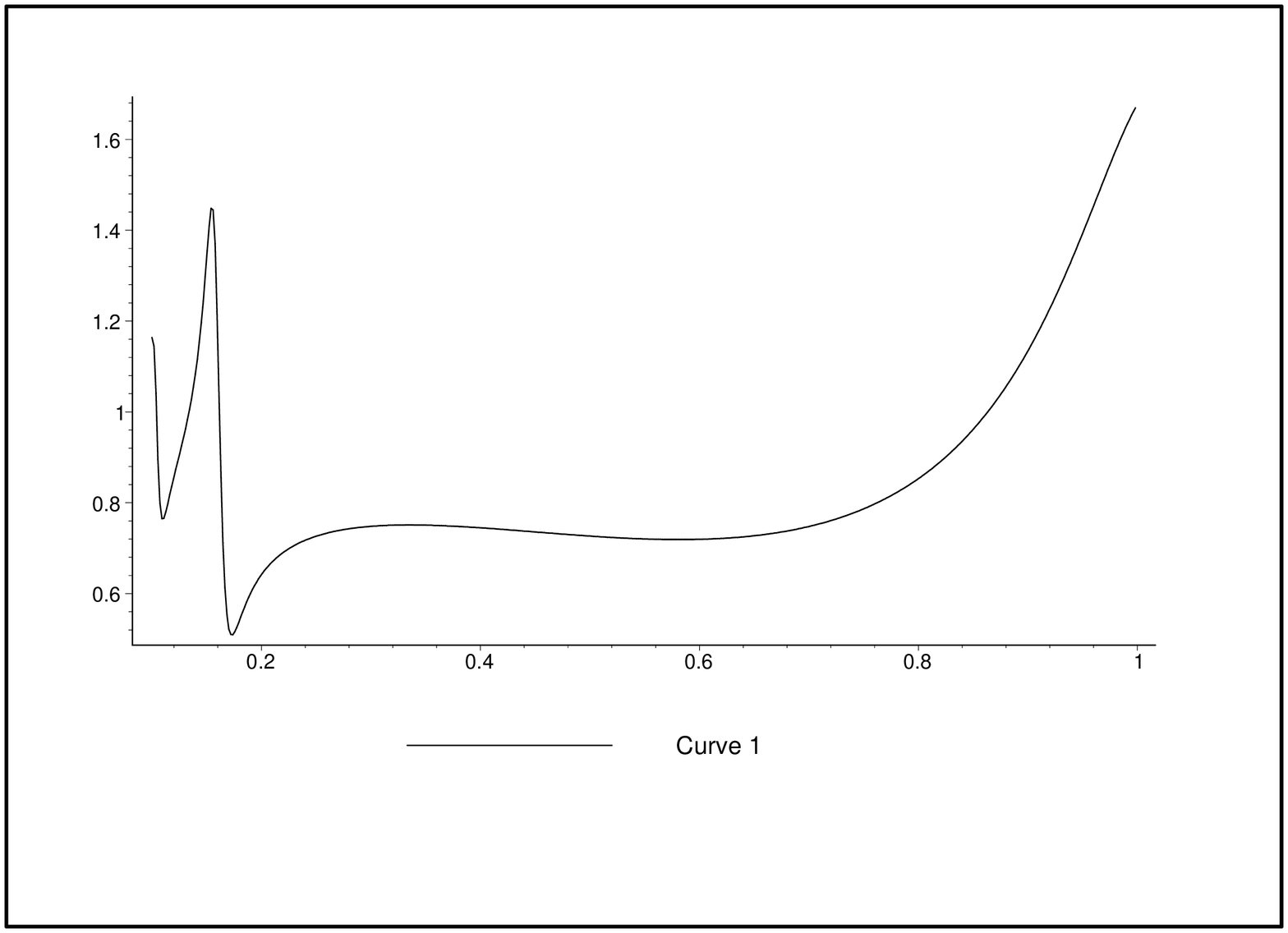}
\caption{The dependence of $\frac{P_{T}^{1/2}}{4\protect\sqrt{\protect%
\epsilon }P_{S}^{1/2}}$ is plotted against $\protect\sigma $. It approaches
the standard result of unity as $\protect\sigma $ goes to zero. $C_{-}$ and $%
D_{-}$ are set to zero.}
\label{fig6}
\end{figure}

Fig.6 shows how the tensor to scalar perturbations ratio varies as a
function of $\sigma $ when $C_{-}=D_{-}=0$. For small values of $\sigma $,
it oscillates about its standard value, $\epsilon $. For intermediate values
of $\sigma $ it remains almost constant on a value that is less than its
standard result. Although the tensor and scalar fluctuations both decrease
as $\sigma $ increases, their ratio gradually increases by increasing $%
\sigma $ and even becomes larger than its standard value. This means that
the tensor fluctuations decrease more slowly than do the scalar fluctuations.

We can derive some qualitative features of the same study for power-law
backgrounds from what we derived in near-de Sitter space. At the beginning
of inflation the expansion is faster than it is at the end of inflation and
so the Hubble parameter is larger. Hence the effect is much more profound
for modes that leave the horizon at that time. For such modes, the ratio
will be much more distorted from standard predictions. We plan
to return to this problem in greater detail \cite{Amjad4}.

Now we assume that $S_{S}^{(1)}$ and $S_{T}^{(2)}$ describes the
behavior of scalar and tensor perturbation. In near-De-sitter and
power-law backgrounds the equation derived from $S_{S}^{(1)}$ and
$S_{T}^{(2)}$ are the same as the ones derived from $S_{T}^{(1)}$
and $S_{S}^{(2)}$ respectively. Therefore the ratio
$\frac{A_{T}^2}{\epsilon A_{S}^2}$ will be reversed. Fig.7 shows
that how the ratio of tensor to scalar perturbations varies as a
function of $\sigma$. The same oscilationary behavior in vicinity
of $\sigma=0$ has repeated. However in this case the tensor/scalar
ratio decreases as the ratio of minimal length approaches the
Hubble length during the inflation. This mechanism might be used
to dampen the contribution of tensor amplitudes to the anisotropy
of the microwave background radiation.
\begin{figure}[t]
\includegraphics[angle=360, width=0.50\textwidth,
height=0.3\textheight]{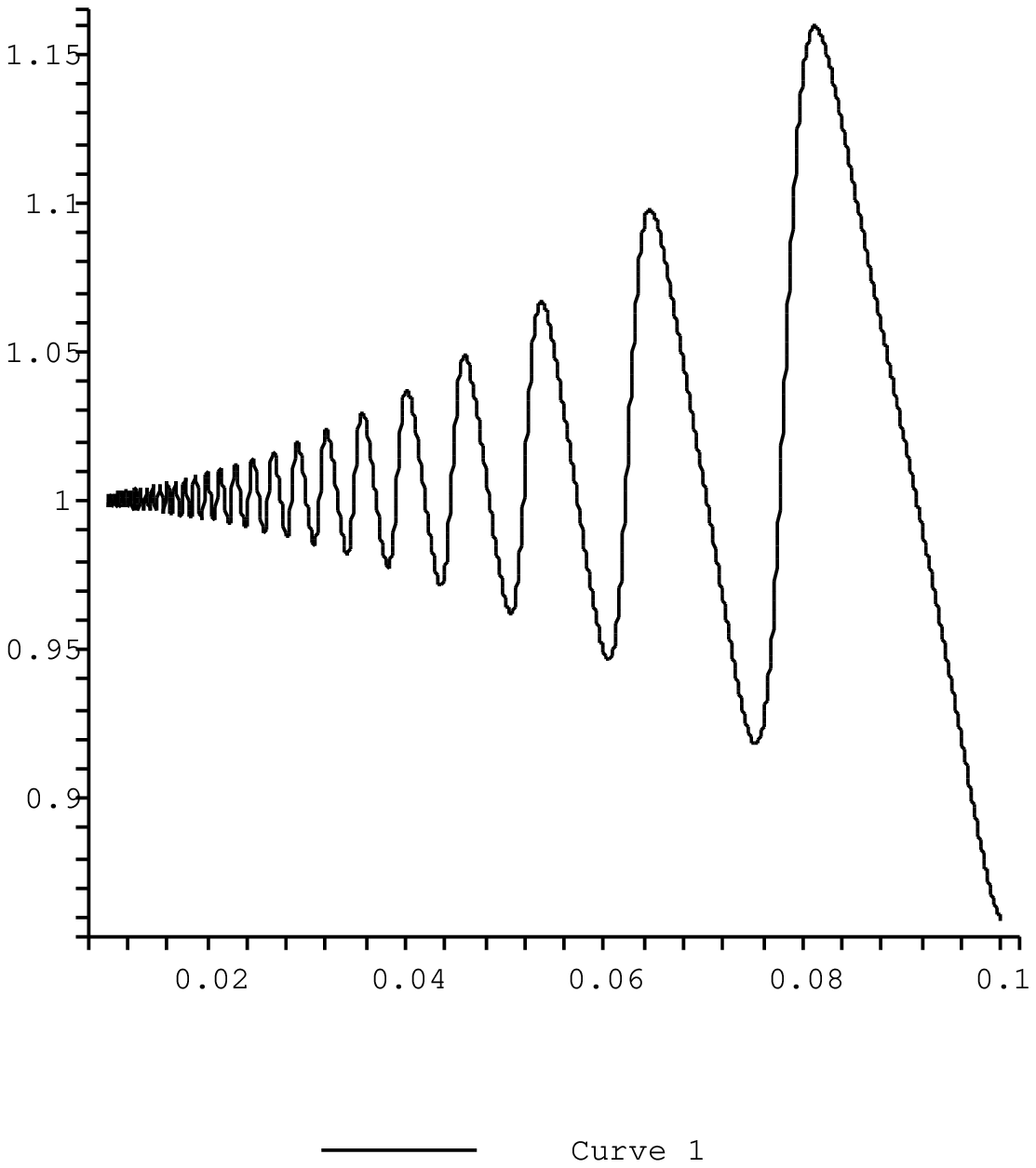}
\includegraphics[angle=360,
width=0.50\textwidth, height=0.3\textheight]{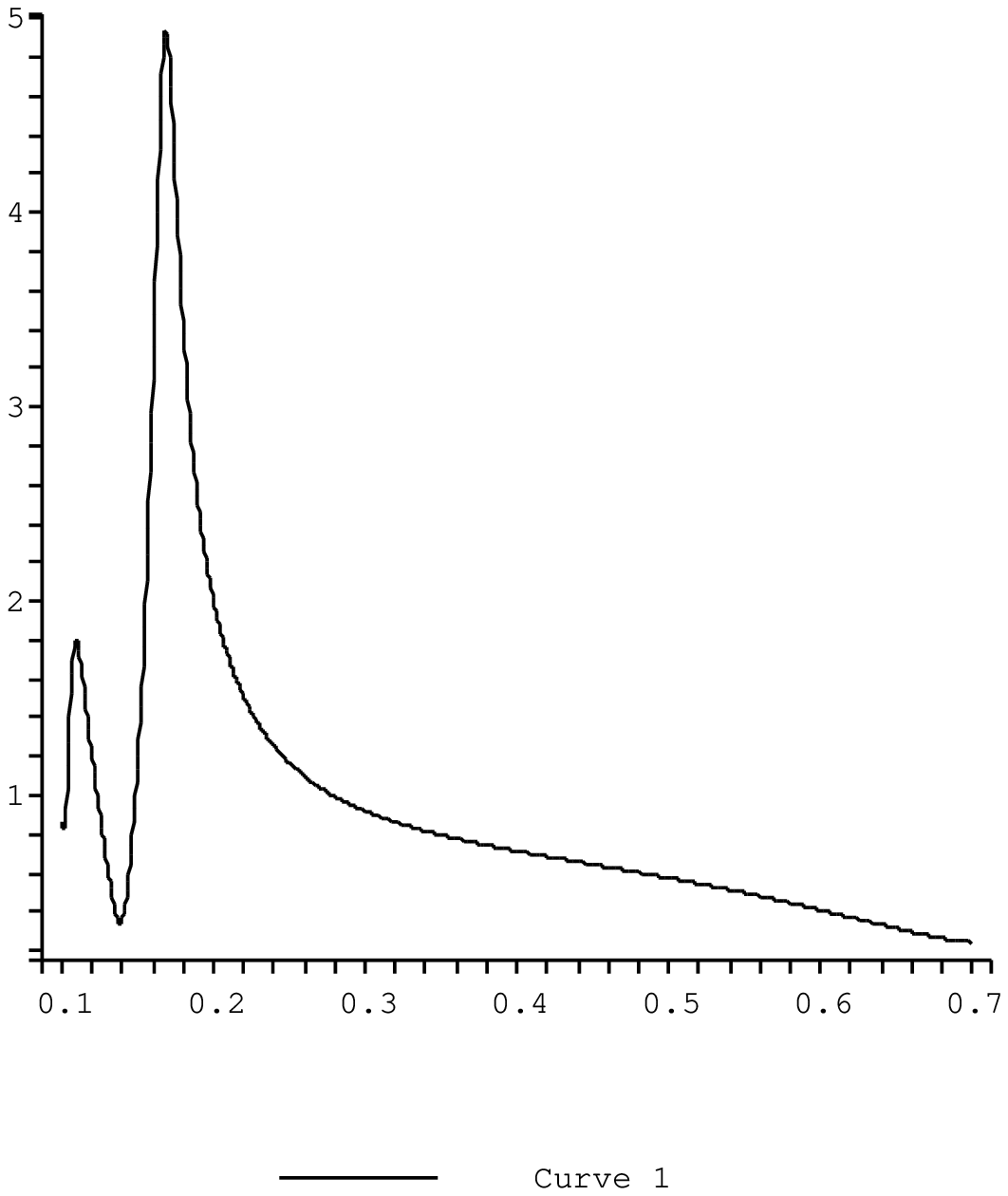}
\caption{The dependence of $\frac{P_{T}^{1/2}}{4\protect\sqrt{\protect%
\epsilon }P_{S}^{1/2}}$ is plotted against $\protect\sigma $. It approaches
the standard result of unity as $\protect\sigma $ goes to zero. $C_{-}$ and $%
D_{-}$ are set to zero. It is assumed that $S_{S}^{(1)}$ and $S_{T}^{(2)}$
describes the behavior of scalar and tensor perturbation, respectively.}
\label{fig7}
\end{figure}

\section{Conclusion}

Both tensor and scalar perturbations are responsible for the anisotropy of
the CMB. Knowledge of the ratio of tensor/scalar perturbations provides an
important constraint on related cosmological parameters.

We have investigated the implications of implementing a minimal
length hypothesis from the generalized uncertainty principle
(\ref{1}) for the tensor/scalar ratio $r$ in inflationary
scenarios. Specifically, we have studied how an ambiguity
generically present in this hypothesis \cite{Amjad1} leads to
different conclusions about how and whether transplanckian physics
alters $r$. In two of the cases the ratio remains constant, unless
the background deviates from a power-law expansion during
inflation. In the other two cases, the ratio is modified even in a
simple de Sitter or power-law background. We also found the
dependence of the ratio on the minimal length for the
near-de-Sitter background in these two cases.

The tensor fluctuations are expected to contribute to the CMB's
$B$ -polarization. This effect may be observable with the upcoming
PLANCK satellite. One can then differentiate the contribution of
tensor fluctuations from scalar ones to check the above scenario.

\bigskip

\section*{Appendix}

The scalar gauge invariant parameter, $u$, is proportional to $\Re
$, the intrinsic curvature perturbations of the spatial
hypersurface through a factor $z$ (\ref{z}) which equivalently can
be defined as:
\begin{equation}
z=\frac{a\dot{\phi _{0}}}{H}  \label{90}
\end{equation}%
where $H$ is the Hubble parameter, $\dot{a}/a$ (dot denotes differentiation
with respect to the physical time). So one obtains:
\begin{equation}
\frac{z^{\prime }}{z}=\frac{a^{\prime }}{a}+\frac{\dot{\phi _{0}}^{\prime }}{%
\dot{\phi _{0}}}-\frac{H^{\prime }}{H}.  \label{91}
\end{equation}%
$\dot{\phi _{0}}^{\prime }/\dot{\phi _{0}}$ can be written as $a\ddot{\phi
_{0}}/\dot{\phi _{0}}$. Using the definition of $\eta $
\begin{equation}
\eta (\phi )\equiv -\frac{\ddot{\phi_{0}}}{H\dot{\phi_{0}}}=\frac{m_{Pl}^{2}}{4\pi }%
\left( \frac{H_{\phi \phi }}{H}\right) =\epsilon -\frac{m_{Pl}{\epsilon }%
_{\phi }}{\sqrt{16\pi \epsilon }},  \label{eta}
\end{equation}%
this can be written in terms of the slow roll parameters:
\begin{equation}
\frac{\dot{\phi _{0}}^{\prime }}{\dot{\phi _{0}}}=-\eta \frac{a^{\prime }}{a}%
.  \label{92}
\end{equation}%
$H^{\prime }/H$ is equal to $a\dot{\phi _{0}}H_{\phi }/H$. Choosing the
convention that $\dot{\phi _{0}}>0$, from Eq. (\ref{eps}) one derives:
\begin{equation}
\frac{H_{\phi }}{H}=-\frac{2\sqrt{\pi \epsilon }}{m_{Pl}}
\label{93}
\end{equation}%
and
\begin{equation}
\dot{\phi _{0}}^{2}=\frac{2\epsilon V}{3-\epsilon }.  \label{94}
\end{equation}%
The inflaton energy density is $\frac{\dot{\phi _{0}}^{2}}{2}+V$ and the
first Friedmann equation
\begin{equation}
H^{2}=\frac{8\pi }{3m_{Pl}^{2}}(\frac{\dot{\phi _{0}}^{2}}{2}+V),
\label{95}
\end{equation}%
combined with (\ref{94}), yields:
\begin{equation}
V=\frac{m_{Pl}^{2}H^{2}(3-\epsilon )}{8\pi }.  \label{96}
\end{equation}%
From Equations (\ref{94})and (\ref{96}) one concludes:
\begin{equation}
\dot{\phi _{0}}=\frac{m_{Pl}H}{2}\sqrt{\frac{\epsilon }{\pi }}.
\label{97}
\end{equation}%
Using the above equation one obtains:
\begin{equation}
\frac{H^{\prime }}{H}=-\epsilon \frac{a^{\prime }}{a}.  \label{98}
\end{equation}%
Inserting equations (\ref{98}) \& (\ref{92}) back into
eq.(\ref{91}), we obtain the following expansion for $z^{\prime
}/z$ in terms of the slow roll parameters:
\begin{equation}
\frac{z^{\prime }}{z}=\frac{a^{\prime }}{a}(1+\epsilon -\eta ).  \label{99}
\end{equation}%
In power-law and near-De-sitter space $\epsilon =\eta $ and so $z^{\prime
}/z=a^{\prime }/a$.

\section*{Acknowledgments}

The authors are thankful to W. H. Kinney, R. H. Brandenberger, R. Easther
and A. Kempf for helpful discussions. This work was supported by the Natural
Sciences \& Engineering Research Council of Canada.

\end{document}